\documentclass[aps,prl,twocolumn,floatfix, showpacs]{revtex4}
\usepackage{graphicx}
\usepackage{dcolumn}
\usepackage{bm}

\newcommand{\be}{\begin{equation}}
\newcommand{\ee}{\end{equation}}

\begin{document}

\title{Search for the ``ultimate state" in turbulent Rayleigh-B\'enard convection}
\author{Denis Funfschilling$^1$}
\author{Eberhard Bodenschatz$^2$}
\author{Guenter Ahlers$^3$}
\affiliation{$^1$LSGC CNRS - GROUPE ENSIC, BP 451, 54001 Nancy Cedex, France\\
$^2$Max Planck Institute for Dynamics and Self-Organization, Am Fassberg 17, D-37077 Goettingen, Germany\\
$^3$Department of Physics, University of California, Santa Barbara, CA 93106, USA}
\date{\today}

\begin{abstract}

Measurements of the Nusselt number $Nu$ and of temperature variations $\Delta T_b$ in the bulk fluid are reported for turbulent Rayleigh-B\'enard convection of a cylindrical sample. They cover the Rayleigh-number range $10^{9} \alt Ra \alt 3\times 10^{14}$ using He (Prandtl number $Pr = 0.67$), N$_2$ ($Pr = 0.72$) and SF$_6$ ($Pr = 0.79$ to 0.84) at pressures up to 15 bars and near-ambient temperatures. The sample had a height $L=2.24$m and diameter $D = 1.12$m and was located in a new High-Pressure Convection Facility (HPCF) at the Max Planck Institute for Dynamics and Self-Organization in G\"ottingen, Germany. The data do not show the transition to an ``ultimate regime" reported by Chavanne et al. and are consistent with the measurements of Niemela et al.

\end{abstract}

\pacs{47.27.te, 47.27.-i, 47.55.P-}

\maketitle

Turbulent convection in a fluid heated from below (Rayleigh-B\'enard convection or RBC) \cite{AGL09} plays a major role in numerous natural processes. It occurs in Earth's outer core \cite{CO94,GCHR99}, atmosphere \cite{DDSC00,HMF01}, and oceans \cite{MS99,Ra00}, and is found in the outer layer of the sun \cite{CEW03} and in giant planets \cite{Bu94}.  The intensity of the driving  by the thermal gradients usually is expressed by the dimensionless temperature difference known as the Rayleigh number $Ra$ (to be defined explicitly below). Another important dimensionless parameter is the ratio of viscous dissipation to thermal dissipation known as the Prandtl number $Pr$. The natural phenomena mentioned above generally involve $Ra \agt 10^{20}$ and a wide range of $Pr$ (see, for instance, Ref. \cite{SD01}), whereas measurements in Earth-bound laboratories, with a few exceptions to be discussed below, had been limited to $Ra \alt 10^{12}$.\cite{AGL09} 

\begin{figure}
\includegraphics[width=3in]{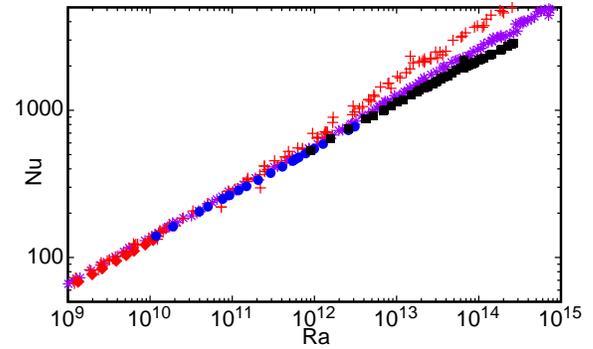}
\caption{The Nusselt number $Nu$ as a function of the Rayleigh number $Ra$ on logarithmic scales.  Plusses (red online): from Chavanne et al. \cite{CCCCH01}. Stars (purple online): from Niemela et al. \cite{NSSD00} after a re-analysis reported in Ref.~\cite{NS06b}. Solid diamonds (red online): this work, He. Solid circles (blue online): this work, N$_2$. Solid squares (black online): this work, SF$_6$.}
\label{fig:N_of_R}
\end{figure}

Of particular interest has been the global heat transport by the turbulent system, as expressed by the Nusselt number $Nu$ (the ratio of the effective conductivity $\lambda_{eff}$ of the convecting system to the conductivity $\lambda$ of the quiescent fluid). Extrapolations of $Nu(Ra,Pr)$  to the geo- or astro-physically relevant ranges are in question because the basic physics involved in the turbulent flow is expected to change at some $Pr$-dependent $Ra = Ra^*(Pr)$. For $Pr \simeq 1$ $Ra^*$ is estimated to be near  $3\times 10^{14}$, and it is expected to increase with $Pr $ approximately as $Pr^{0.7}$.\cite{GL02} Unfortunately, on this important issue two nominally equivalent sets of $Nu$ measurements \cite{CCCHCC97}\cite{NSSD00,NSSD00e} in the parameter ranges where $Ra^*$ might be found disagree with each other. One of them, by a group in Grenoble, France \cite{CCCHCC97,CCCCH01} (the ``Grenoble" data) and shown in Fig.~\ref{fig:N_of_R} as plusses, was interpreted as evidence for $Ra^*$ at an unexpectedly low value near $10^{11}$ (this evidence will become more obvious in Fig.~\ref{fig:Nred_of_R} below). The other, by a group in Oregon, \cite{NSSD00,NS06b} (the ``Oregon" data) shown in Fig.~\ref{fig:N_of_R} as stars, did not show any transition all the way up to $Ra \simeq 10^{17}$. Both experiments were done at approximately 5 K using helium near its critical point. Their relationships to $Ra^*$ are somewhat  uncertain. Because of the proximity of the critical point $Pr$ increased significantly as $Ra$ increased beyond about $10^{12}$ and thus the expected value of $Ra^*$ increased. 

Here we report new measurements, made at ambient temperatures using compresses helium (He), nitrogen (N$_2$), and sulfur hexafluoride (SF$_6$), over the range $10^{9} \alt Ra \alt 3\times 10^{14}$. They are shown as solid symbols in Fig.~\ref{fig:N_of_R}. Our data fall into the narrow range $0.67 \alt Pr \alt 0.84$ characteristic of classical gases where theoretical estimates \cite{GL02} give $Ra^* \simeq 2\times 10^{14}$, albeit with a considerable uncertainty. Our measurements do not reveal the anticipated transition to an ultimate regime; they are roughly consistent with the Oregon data but are inconsistent with the Grenoble data. It remains unclear to us what caused the transition revealed by the Grenoble results.

For modest, laboratory-accessible, $Ra$ the heat transport is controlled essentially by thin thermal boundary layers (BLs) just above (below) the bottom (top) plate of the convection cell. For $Ra > Ra^*$ it is expected that a large-scale circulation (LSC) in the sample interior will apply sufficient shear to the BLs to cause them to undergo a turbulent transition. This transition was predicted to occur when the shear Reynolds number $Re_s$, based on the BL thickness, exceeds a critical value $Re_s^* \simeq 420$.\cite{GL02} It is expected to lead to a new turbulent convective state in the sample interior that was analyzed theoretically by Kraichnan \cite{Kr62} and that is often referred to as the "ultimate state" because it should prevail asymptotically up to infinite $Ra$. Clearly the experimental discovery and study of the ultimate state is one of the major challenges to experiment in this field, both because of its fundamental interest and because of its geo/astrophysical relevance. This is the reason why understanding the difference between the Oregon and the Grenoble data is a major issue in the field of turbulent convection. 

Before proceeding to a discussion of our results, it is useful to re-examine the prior large-$Ra$ measurements in more detail. Over 30 years ago it was recognized that the study of RBC using low-temperature helium has unique advantages because of the unusual properties of this fluid. \cite{Ah74,Ah75,Th75} This was exploited in early experiments by a group in Chicago who reached $Ra \simeq 6\times 10^{12}$ \cite{CGHKLTWZZ89} using a cylindrical sample of aspect ratio $\Gamma \equiv D/L = 0.5$ ($D$ is the diameter and  $L$ the height) by approaching the critical region of Helium where $Ra$ tends to become large; but that work  did not reveal any transition in the heat transport. Later the Grenoble group \cite{CCCCCH96} carried out similar measurements, also using a cylinder with $\Gamma = 0.5$, and tentatively identified a transition at $Ra^* \simeq 10^{11}$. 
Below that point $Nu$ roughly followed an effective power law $Nu = Ra^{\gamma_{eff}}$, with $\gamma_{eff} \simeq 0.3$. Above $Ra^*$ $\gamma_{eff}$  approached values near 0.4 and was heading toward the asymptotic value 0.5
 predicted by Kraichnan. Soon thereafter the same group published a Physical Review Letter entitled ``Observation of the ultimate regime in {{Rayleigh-B\'enard}} convection". \cite{CCCHCC97} A number of other papers by this group followed with more detail and data up to
  $Ra \simeq 10^{15}$, and we show the results reported in Ref.~\cite{CCCCH01} in 
  Fig.~\ref{fig:N_of_R} as plusses. It remained unexplained why the Chicago group 
  did not find the transition seen by the Grenoble group.

A few years later the Oregon group extended the helium measurements to theretofore unprecedented values of $Ra$ as large as $10^{17}$ by building a cylindrical convection cell with a height $L = 1.0$ m and a diameter $D=0.5$ m and operating near the critical point of helium.\cite{NSSD00} Their data\cite{NS06b} are shown as stars in Fig.~\ref{fig:N_of_R} and do not reveal any transition. They can be described over the wide range of $Ra$ by a power law with $\gamma_{eff} \simeq 0.32$ with no significant tendency for $\gamma_{eff}$ to increase with $Ra$.

We note that the Chicago, Grenoble, and Oregon experiments were conducted at cryogenic temperatures where experimental difficulties are quite severe and where on average $Pr$ tends to increased with $Ra$. It was very desirable to carry out an experiment at near-ambient temperatures using more conventional, and above all different, experimental techniques and fluids with essentially constant $Pr$ that could also achieve the large-$Ra$ values deemed necessary to reach $Ra^*$. To satisfy those criteria, we used a very large pressure vessel at the Max Planck Institute for Dynamics and Self-Organization in G\"ottingen, Germany. It is a cylinder of diameter 2.5 m and length 5.5 m, 
 with its axis horizontal, and with a turret above it that extends the height to 4 m over a diameter of 1.5 m. Because of its suggestive shape, this vessel has become known as the ``Uboot of G\"ottingen". 
It has an approximate volume of 25 m$^3$ and can be used up to pressures of 15 bars.  In the section containing the turret we placed a RBC sample-cell with $L = 2.24$ m and $D = 1.12$ m (the  ``High Pressure Convection Facility" or HPCF), yielding $\Gamma = 0.500$. It had top and bottom plates made of Aluminum and a Plexiglas side wall of thickness 9 mm. The top plate extended from the top a short distance into the side wall; it was water cooled with flow through quadruple spiral channels. The bottom plate was a composite consisting of two aluminum plates with a Plexiglas plate of thickness 5 mm between them, all glued together with Stycast 1266 epoxy. The composite extended from the bottom into the side wall. Joule heating was applied with a heater that was uniformly distributed over the bottom surface of the bottom plate. Aside from that total heat current $Q$, the heat current $Q_s$ entering the sample could be inferred from the temperature difference across the composite. \cite{Ma54,KH81} Under most conditions $Q_s$ was equal to $Q$, indicating no significant parasitic heat losses from the bottom-plate heater. In all cases $Q_s$ was used to compute $Nu$. Various thermal shields, regulated at appropriate temperatures to prevent heat losses, had been installed as described for instance in Ref.~\cite{BFNA05}. All spaces outside of the sample and up to a diameter of 1.4 m were occupied by low-density open-pore foam. Estimates indicate that side-wall heat-losses were negligible. Each data point was derived from a time series of temperature readings at time intervals of about 5 sec and spanning about one day. The first half of each series was discarded to avoid transients and the remainder was time-averaged before computing $Ra$ and $Nu$. The data were corrected for the finite plate conductivity \cite{Ve04,BFNA05}; this correction generally was less than a few percent. 

The Rayleigh number is given by $Ra = \beta g L^3 \Delta T/(\kappa \nu)$, and the Prandtl number is $Pr = \nu/\kappa$. Here the thermal expansion coefficient $\beta$,  the thermal diffusivity $\kappa$, and the kinematic viscosity $\nu$ were evaluated at the mean temperature $T_m = (T_t+T_b)/2$ ($T_t$ and $T_b$ are the temperatures at the top and  bottom of the fluid respectively). The gravitational acceleration is $g$, and $\Delta T = T_b-T_t$. The Nusselt number is $Nu = Q_s L / (A\Delta T \lambda)$ where $A$ is the sample cross-sectional area and $\lambda$ is the thermal conductivity. The fluid properties were compiled previously \cite{LA97} from numerous  papers in the literature.

\begin{figure}
\includegraphics[width=2.5in]{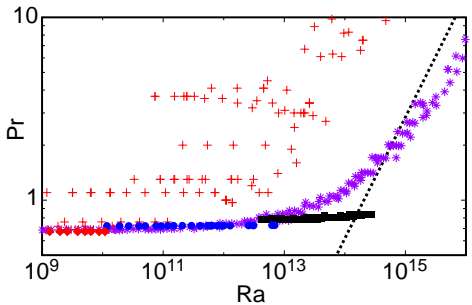}
\caption{The Prandtl number as a function of the Rayleigh number corresponding to the various data sets. The symbols are as in Fig.~\ref{fig:N_of_R}. The dotted line is the estimate of $Pr(Ra^*)$ from Ref.~\cite{GL02} for $\Gamma = 1$.}
\label{fig:Pr_of_Ra}
\end{figure}

Measurements of $Nu(Ra)$ were made with He ($Pr = 0.67$), N$_2$ ($Pr = 0.72$), and two separate gas fillings (run 1 and run 2) of SF$_6$ ($Pr = 0.79$ to 84). The data span the  ranges $10^{9} \leq Ra \leq 3\times 10^{14}$.
The results are shown in Fig.~\ref{fig:N_of_R} as solid symbols. Figure~\ref{fig:Pr_of_Ra} shows the points in the $Ra-Pr$ parameter space where the various data were taken. 
Even though at large $Ra$ the Prandtl number is much lower (and thus $Ra^*$ is much lower) than it is for the Oregon or Grenoble data, our data in Fig.~\ref{fig:N_of_R} reveal no transition to an ultimate regime. The dotted line in Fig.~\ref{fig:Pr_of_Ra} is an estimate of $Ra^*(Pr)$ for cylindrical samples  with $\Gamma = 1$ which may serve as a rough guide of where to expect the transition also for $\Gamma = 0.5$. We see that the Oregon data and our new results exceed that estimate by about the same amount even though the Oregon data extend to much larger $Ra$. The Grenoble data, on the other hand, do not come near the estimate; thus it seems unlikely that the transition near $Ra = 10^{11}$ which they reveal  is associated with a shear-induced turbulence transition in the BLs.

\begin{figure}
\includegraphics[width=3in]{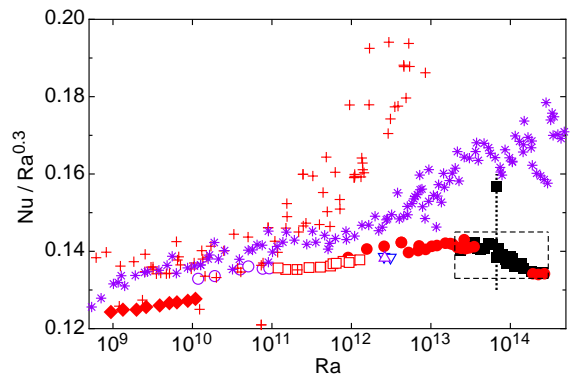}
\caption{The reduced Nusselt number $Nu/Ra^{0.3}$ as a function of the Rayleigh number $Ra$. Plusses (red online): from Chavanne et al. \cite{CCCCH01}. Stars purple online): from Niemela et al. \cite{NSSD00} after a re-analysis reported in Ref.~\cite{NS06b}. Solid diamonds (red online): this work, He, $P = 4.3$ bars. Solid squares (black online): SF$_6$, run 1. Solid circles (red online): SF$_6$, run 2. Open symbols: This work, N$_2$. Circles (purple online): $P = 2$ bars.  Squares (red online): $P = 6$ bars.  Down pointing triangles (blue online): $P = 10$ bars.  Up pointing triangles (purple online): $P = 15$ bars.The vertical dotted line shows the location of a transition, see text. The area in the dashed rectangle is shown enlarged in Fig.~\ref{fig:hires}a.}
\label{fig:Nred_of_R}
\end{figure}

\begin{figure}
\includegraphics[width=2.75in]{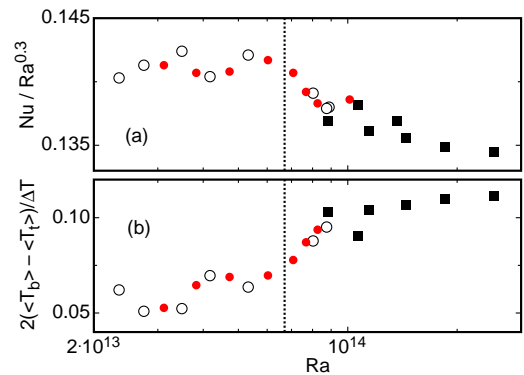}
\caption{Details from SF$_6$, run 1, of (a) the reduced Nusselt number $Nu/Ra^{0.3}$  and (b) the temperature drop $2(\langle T_b\rangle - \langle T_t\rangle)/\Delta T$ in the bulk near the side wall. Open (solid) circles: $P \simeq 8$ bars, decreasing (increasing) pressure. Solid squares: $15 \agt P \agt 9$ bars. The vertical dotted lines show the location of a transition, see text.}
\label{fig:hires}
\end{figure}

In Fig.~\ref{fig:Nred_of_R} we show the data for $Nu$ in more detail by dividing out the approximate $Ra$ dependence $Ra^{0.3}$. For $Ra \alt 10^{11}$ there is general consistency of the dependence of $Nu$ on $Ra$ between all three data sets. Our He data are about 6\% below the Oregon and Grenoble data and about 4\% below our own N$_2$ data. Possible causes for these differences may be systematic errors in the calibration of the thermal resistance of the bottom-plate composite, systematic errors in the thermodynamic and transport properties used to compute $Ra$ and  $Nu$ for the various fluids, as well as an expected increase of $Nu$ with $Pr$ at constant $Ra$ for $Ra \alt 1$. For this ``reduced" Nusselt number the break in the Grenoble data at $Ra \simeq 10^{11}$ is more apparent than it was in Fig.~\ref{fig:N_of_R}. Clearly this transition to a significantly larger effective exponent, near $\gamma_{eff} = 0.39$, is not present in the Oregon data; nor can it be found in the G\"ottingen data. In general our data are slightly lower that the Oregon data, and this difference  increases as $Ra$ increases. In part this may again be due to systematic errors in the fluid properties that were used in the analyses, especially near the critical point of helium; but at the larger $Ra$ a dependence of the Oregon $Nu$ data on $Pr$ may also be contributing. As illustrated in Fig.~\ref{fig:Pr_of_Ra}, $Pr$ starts to increase for the Oregon data as $Ra$ exceeds about $10^{12}$, whereas for our data $Pr$ is nearly independent of $Ra$.

A notable feature of the Goettingen data is a sudden change of the dependence of $Nu$ upon $Ra$ near $Ra = 6.7\times 10^{13}$.  Hardly noticeable in Fig.~\ref{fig:N_of_R}, it becomes apparent with the much higher resolution of Fig.~\ref{fig:Nred_of_R}
where it is indicated by a vertical dotted line. One data point, per chance taken precisely at that transition, yielded a $Nu$ value 12\% higher than the neighboring ones (see the isolated solid square in Fig.~\ref{fig:Nred_of_R}).  To look at this feature more closely, we show  in Fig.~\ref{fig:hires}a the data from run 1, SF$_6$, in the dashed rectangle of Fig.~\ref{fig:Nred_of_R} on an enlarged scale. There the circles are for $P \simeq 8$ bars and the squares are for higher pressures. The open (solid) circles were taken with decreasing (increasing) $Ra$ (and thus $\Delta T$). Again the transition is indicated by the vertical dotted line. We believe that it reflects a change in the structure of the LSC, perhaps from a single roll to a two-roll structure as found earlier in direct numerical simulations \cite{VC03,SV06} and experiments \cite{XX07}, but more work is required to be more specific. 

An interesting and instructive feature of this system is the vertical temperature drop $\Delta T_b$ across the bulk of the fluid.\cite{BA07_EPL} Whereas the temperature drop $\Delta T_{BL}$ across the two boundary layers dominates $\Delta T = \Delta T_{BL} + \Delta T_b$, values of order 0.1 have been found for $\Delta T_b/\Delta T$ in experiments with water,  $Ra$ near $10^9$, and near the side wall where plumes travel from one plate to the other (any temperature drop along the sample axis where plumes are scarce was found to be considerably smaller). In the range $Ra < Ra^*$ the gradient in the bulk is expected to remain small; but as $Ra^*$ is exceeded, the BLs are expected to disappear and thus most of the temperature drop should occur across the bulk. Thus we anticipate a dramatic change of $\Delta T_b$ as $Ra^*$ is exceeded.
With this in mind, the azimuthal and temporal mean temperatures $\langle T_b\rangle$ and $\langle T_t\rangle$ along the side wall at the vertical positions $-L/4$ and $L/4$ respectively  were determined with thermometers imbedded in the side walls as described for instance in Ref.~\cite{BA07_EPL}.  Those data yield the estimates $\Delta T_b \simeq 2(\langle T_b\rangle - \langle T_t\rangle)$ shown in Fig.~\ref{fig:hires}b. There is some potentially interesting structure in the $Ra$ dependence of this quantity, especially near the transition at $Ra = 6.8\times 10^{13}$ which is indicated by the vertical dotted line; but the overall value of $\Delta T_b/\Delta T$ remains modest at all $Ra$ and is comparable to similar results reported elsewhere.\cite{BA07_EPL} The results imply that $\Delta T_{BL} >> \Delta T_b$, consistent with $Ra < Ra^*$, for our entire parameter range.

In this Letter we report new measurements of the Nusselt number, and of the bulk temperature gradient near the side wall, for turbulent Rayleigh-B\'enard convection in a cylindrical sample of aspect ratio 0.5 over the Rayleigh-number range $10^9 \alt Ra \alt 3\times 10^{14}$ and for Prandtl numbers close to 0.8. The results do not reveal any evidence for a transition to the ``ultimate" regime expected in the large-$Ra$ limit.

We are very greatful to the Max-Planck-Society and the Volkswagen Stiftung, whose generous support made the establishment of the facility and the experiments possible. The work of G.A. was supported in part by the U.S National Science Foundation through Grant DMR07-02111. We are very grateful to Artur Kubitzek and Andreas Renner for their enthusiastic technical support and to Holger Nobach for his role in developing the SF$_6$ system.
\vskip -0.3in


\begin{thebibliography}{31}
\expandafter\ifx\csname natexlab\endcsname\relax\def\natexlab#1{#1}\fi
\expandafter\ifx\csname bibnamefont\endcsname\relax
  \def\bibnamefont#1{#1}\fi
\expandafter\ifx\csname bibfnamefont\endcsname\relax
  \def\bibfnamefont#1{#1}\fi
\expandafter\ifx\csname citenamefont\endcsname\relax
  \def\citenamefont#1{#1}\fi
\expandafter\ifx\csname url\endcsname\relax
  \def\url#1{\texttt{#1}}\fi
\expandafter\ifx\csname urlprefix\endcsname\relax\def\urlprefix{URL }\fi
\providecommand{\bibinfo}[2]{#2}
\providecommand{\eprint}[2][]{\url{#2}}

\bibitem[{\citenamefont{Ahlers et~al.}(2009)\citenamefont{Ahlers, Grossmann,
  and Lohse}}]{AGL09}
\bibinfo{author}{\bibfnamefont{G.}~\bibnamefont{Ahlers}},
  \bibinfo{author}{\bibfnamefont{S.}~\bibnamefont{Grossmann}},
  \bibnamefont{and} \bibinfo{author}{\bibfnamefont{D.}~\bibnamefont{Lohse}},
  \bibinfo{journal}{Rev. Mod. Phys.} \textbf{\bibinfo{volume}{81}},
  \bibinfo{pages}{in print} (\bibinfo{year}{2009}).

\bibitem[{\citenamefont{Cardin and Olson}(1994)}]{CO94}
\bibinfo{author}{\bibfnamefont{P.}~\bibnamefont{Cardin}} \bibnamefont{and}
  \bibinfo{author}{\bibfnamefont{P.}~\bibnamefont{Olson}},
  \bibinfo{journal}{Phys. of the Earth and Planetary Interiors}
  \textbf{\bibinfo{volume}{82}}, \bibinfo{pages}{235} (\bibinfo{year}{1994}).

\bibitem[{\citenamefont{Glatzmaier et~al.}(1999)\citenamefont{Glatzmaier, Coe,
  Hongre, and Roberts}}]{GCHR99}
\bibinfo{author}{\bibfnamefont{G.}~\bibnamefont{Glatzmaier}},
  \bibinfo{author}{\bibfnamefont{R.}~\bibnamefont{Coe}},
  \bibinfo{author}{\bibfnamefont{L.}~\bibnamefont{Hongre}}, \bibnamefont{and}
  \bibinfo{author}{\bibfnamefont{P.}~\bibnamefont{Roberts}},
  \bibinfo{journal}{Nature(London)} \textbf{\bibinfo{volume}{401}},
  \bibinfo{pages}{885} (\bibinfo{year}{1999}).

\bibitem[{\citenamefont{van Doorn et~al.}(2000)\citenamefont{van Doorn, Dhruva,
  Sreenivasan, and Cassella}}]{DDSC00}
\bibinfo{author}{\bibfnamefont{E.}~\bibnamefont{van Doorn}},
  \bibinfo{author}{\bibfnamefont{B.}~\bibnamefont{Dhruva}},
  \bibinfo{author}{\bibfnamefont{K.~R.} \bibnamefont{Sreenivasan}},
  \bibnamefont{and} \bibinfo{author}{\bibfnamefont{V.}~\bibnamefont{Cassella}},
  \bibinfo{journal}{Phys. Fluids} \textbf{\bibinfo{volume}{12}},
  \bibinfo{pages}{1529} (\bibinfo{year}{2000}).

\bibitem[{\citenamefont{Hartmann et~al.}(2001)\citenamefont{Hartmann, Moy, and
  Fu}}]{HMF01}
\bibinfo{author}{\bibfnamefont{D.~L.} \bibnamefont{Hartmann}},
  \bibinfo{author}{\bibfnamefont{L.~A.} \bibnamefont{Moy}}, \bibnamefont{and}
  \bibinfo{author}{\bibfnamefont{Q.}~\bibnamefont{Fu}}, \bibinfo{journal}{J.
  Climate} \textbf{\bibinfo{volume}{14}}, \bibinfo{pages}{4495}
  (\bibinfo{year}{2001}).

\bibitem[{\citenamefont{Marshall and Schott}(1999)}]{MS99}
\bibinfo{author}{\bibfnamefont{J.}~\bibnamefont{Marshall}} \bibnamefont{and}
  \bibinfo{author}{\bibfnamefont{F.}~\bibnamefont{Schott}},
  \bibinfo{journal}{Rev. Geophys.} \textbf{\bibinfo{volume}{37}},
  \bibinfo{pages}{1} (\bibinfo{year}{1999}).

\bibitem[{\citenamefont{Rahmstorf}(2000)}]{Ra00}
\bibinfo{author}{\bibfnamefont{S.}~\bibnamefont{Rahmstorf}},
  \bibinfo{journal}{Climate Change} \textbf{\bibinfo{volume}{46}},
  \bibinfo{pages}{247} (\bibinfo{year}{2000}).

\bibitem[{\citenamefont{Cattaneo et~al.}(2003)\citenamefont{Cattaneo, Emonet,
  and Weiss}}]{CEW03}
\bibinfo{author}{\bibfnamefont{F.}~\bibnamefont{Cattaneo}},
  \bibinfo{author}{\bibfnamefont{T.}~\bibnamefont{Emonet}}, \bibnamefont{and}
  \bibinfo{author}{\bibfnamefont{N.}~\bibnamefont{Weiss}},
  \bibinfo{journal}{Astrophys. J.} \textbf{\bibinfo{volume}{588}},
  \bibinfo{pages}{1183} (\bibinfo{year}{2003}).

\bibitem[{\citenamefont{Busse}(1994)}]{Bu94}
\bibinfo{author}{\bibfnamefont{F.~H.} \bibnamefont{Busse}},
  \bibinfo{journal}{Chaos} \textbf{\bibinfo{volume}{4}}, \bibinfo{pages}{123}
  (\bibinfo{year}{1994}).

\bibitem[{\citenamefont{Sreenivasan and Donnelly}(2001)}]{SD01}
\bibinfo{author}{\bibfnamefont{K.~R.} \bibnamefont{Sreenivasan}}
  \bibnamefont{and} \bibinfo{author}{\bibfnamefont{R.~J.}
  \bibnamefont{Donnelly}}, \bibinfo{journal}{Adv. Appl. Mech.}
  \textbf{\bibinfo{volume}{37}}, \bibinfo{pages}{239} (\bibinfo{year}{2001}).

\bibitem[{\citenamefont{Grossmann and Lohse}(2002)}]{GL02}
\bibinfo{author}{\bibfnamefont{S.}~\bibnamefont{Grossmann}} \bibnamefont{and}
  \bibinfo{author}{\bibfnamefont{D.}~\bibnamefont{Lohse}},
  \bibinfo{journal}{Phys. Rev. E} \textbf{\bibinfo{volume}{66}},
  \bibinfo{pages}{016305} (\bibinfo{year}{2002}).

\bibitem[{\citenamefont{Chavanne et~al.}(1997)\citenamefont{Chavanne, Chilla,
  Castaing, Hebral, Chabaud, and Chaussy}}]{CCCHCC97}
\bibinfo{author}{\bibfnamefont{X.}~\bibnamefont{Chavanne}},
  \bibinfo{author}{\bibfnamefont{F.}~\bibnamefont{Chilla}},
  \bibinfo{author}{\bibfnamefont{B.}~\bibnamefont{Castaing}},
  \bibinfo{author}{\bibfnamefont{B.}~\bibnamefont{Hebral}},
  \bibinfo{author}{\bibfnamefont{B.}~\bibnamefont{Chabaud}}, \bibnamefont{and}
  \bibinfo{author}{\bibfnamefont{J.}~\bibnamefont{Chaussy}},
  \bibinfo{journal}{Phys. Rev. Lett.} \textbf{\bibinfo{volume}{79}},
  \bibinfo{pages}{3648} (\bibinfo{year}{1997}).

\bibitem[{\citenamefont{Niemela
  et~al.}(2000{\natexlab{a}})\citenamefont{Niemela, Skrebek, Sreenivasan, and
  Donnelly}}]{NSSD00}
\bibinfo{author}{\bibfnamefont{J.~J.} \bibnamefont{Niemela}},
  \bibinfo{author}{\bibfnamefont{L.}~\bibnamefont{Skrebek}},
  \bibinfo{author}{\bibfnamefont{K.~R.} \bibnamefont{Sreenivasan}},
  \bibnamefont{and} \bibinfo{author}{\bibfnamefont{R.}~\bibnamefont{Donnelly}},
  \bibinfo{journal}{Nature} \textbf{\bibinfo{volume}{404}},
  \bibinfo{pages}{837} (\bibinfo{year}{2000}{\natexlab{a}}).

\bibitem[{\citenamefont{Niemela
  et~al.}(2000{\natexlab{b}})\citenamefont{Niemela, Skrebek, Sreenivasan, and
  Donnelly}}]{NSSD00e}
\bibinfo{author}{\bibfnamefont{J.~J.} \bibnamefont{Niemela}},
  \bibinfo{author}{\bibfnamefont{L.}~\bibnamefont{Skrebek}},
  \bibinfo{author}{\bibfnamefont{K.~R.} \bibnamefont{Sreenivasan}},
  \bibnamefont{and} \bibinfo{author}{\bibfnamefont{R.}~\bibnamefont{Donnelly}},
  \bibinfo{journal}{Nature} \textbf{\bibinfo{volume}{406}}, \bibinfo{pages}{439
  (erratum)} (\bibinfo{year}{2000}{\natexlab{b}}).

\bibitem[{\citenamefont{Chavanne et~al.}(2001)\citenamefont{Chavanne, Chilla,
  Chabaud, Castaing, and Hebral}}]{CCCCH01}
\bibinfo{author}{\bibfnamefont{X.}~\bibnamefont{Chavanne}},
  \bibinfo{author}{\bibfnamefont{F.}~\bibnamefont{Chilla}},
  \bibinfo{author}{\bibfnamefont{B.}~\bibnamefont{Chabaud}},
  \bibinfo{author}{\bibfnamefont{B.}~\bibnamefont{Castaing}}, \bibnamefont{and}
  \bibinfo{author}{\bibfnamefont{B.}~\bibnamefont{Hebral}},
  \bibinfo{journal}{Phys. Fluids} \textbf{\bibinfo{volume}{13}},
  \bibinfo{pages}{1300} (\bibinfo{year}{2001}).

\bibitem[{\citenamefont{Niemela and Sreenivasan}(2006)}]{NS06b}
\bibinfo{author}{\bibfnamefont{J.~J.} \bibnamefont{Niemela}} \bibnamefont{and}
  \bibinfo{author}{\bibfnamefont{K.~R.} \bibnamefont{Sreenivasan}},
  \bibinfo{journal}{J. Low Temp. Phys.} \textbf{\bibinfo{volume}{143}},
  \bibinfo{pages}{163} (\bibinfo{year}{2006}).

\bibitem[{\citenamefont{Kraichnan}(1962)}]{Kr62}
\bibinfo{author}{\bibfnamefont{R.~H.} \bibnamefont{Kraichnan}},
  \bibinfo{journal}{Phys. Fluids} \textbf{\bibinfo{volume}{5}},
  \bibinfo{pages}{1374} (\bibinfo{year}{1962}).

\bibitem[{\citenamefont{Ahlers}(1974)}]{Ah74}
\bibinfo{author}{\bibfnamefont{G.}~\bibnamefont{Ahlers}},
  \bibinfo{journal}{Phys. Rev. Lett.} \textbf{\bibinfo{volume}{33}},
  \bibinfo{pages}{1185} (\bibinfo{year}{1974}).

\bibitem[{\citenamefont{Threlfall}(1975)}]{Th75}
\bibinfo{author}{\bibfnamefont{D.~C.} \bibnamefont{Threlfall}},
  \bibinfo{journal}{J. Fluid Mech.} \textbf{\bibinfo{volume}{67}},
  \bibinfo{pages}{17} (\bibinfo{year}{1975}).

\bibitem[{\citenamefont{Ahlers}(1975)}]{Ah75}
\bibinfo{author}{\bibfnamefont{G.}~\bibnamefont{Ahlers}}, in
  \emph{\bibinfo{booktitle}{Fluctuations, Instabilities and Phase
  Transitions}}, edited by
  \bibinfo{editor}{\bibfnamefont{T.}~\bibnamefont{Riste}}
  (\bibinfo{publisher}{Plenum}, \bibinfo{address}{New York},
  \bibinfo{year}{1975}), pp. \bibinfo{pages}{181--193}.

\bibitem[{\citenamefont{Castaing et~al.}(1989)\citenamefont{Castaing,
  Gunaratne, Heslot, Kadanoff, Libchaber, Thomae, Wu, Zaleski, and
  Zanetti}}]{CGHKLTWZZ89}
\bibinfo{author}{\bibfnamefont{B.}~\bibnamefont{Castaing}},
  \bibinfo{author}{\bibfnamefont{G.}~\bibnamefont{Gunaratne}},
  \bibinfo{author}{\bibfnamefont{F.}~\bibnamefont{Heslot}},
  \bibinfo{author}{\bibfnamefont{L.}~\bibnamefont{Kadanoff}},
  \bibinfo{author}{\bibfnamefont{A.}~\bibnamefont{Libchaber}},
  \bibinfo{author}{\bibfnamefont{S.}~\bibnamefont{Thomae}},
  \bibinfo{author}{\bibfnamefont{X.~Z.} \bibnamefont{Wu}},
  \bibinfo{author}{\bibfnamefont{S.}~\bibnamefont{Zaleski}}, \bibnamefont{and}
  \bibinfo{author}{\bibfnamefont{G.}~\bibnamefont{Zanetti}},
  \bibinfo{journal}{J. Fluid Mech.} \textbf{\bibinfo{volume}{204}},
  \bibinfo{pages}{1} (\bibinfo{year}{1989}).

\bibitem[{\citenamefont{Chavanne et~al.}(1996)\citenamefont{Chavanne, Chill\'a,
  Chabaud, Castang, Chaussy, and H\'ebral}}]{CCCCCH96}
\bibinfo{author}{\bibfnamefont{X.}~\bibnamefont{Chavanne}},
  \bibinfo{author}{\bibfnamefont{F.}~\bibnamefont{Chill\'a}},
  \bibinfo{author}{\bibfnamefont{B.}~\bibnamefont{Chabaud}},
  \bibinfo{author}{\bibfnamefont{B.}~\bibnamefont{Castang}},
  \bibinfo{author}{\bibfnamefont{J.}~\bibnamefont{Chaussy}}, \bibnamefont{and}
  \bibinfo{author}{\bibfnamefont{B.}~\bibnamefont{H\'ebral}},
  \bibinfo{journal}{J. Low Temp. Phys.} \textbf{\bibinfo{volume}{104}},
  \bibinfo{pages}{109} (\bibinfo{year}{1996}).

\bibitem[{\citenamefont{Malkus}(1954)}]{Ma54}
\bibinfo{author}{\bibfnamefont{M.~V.~R.} \bibnamefont{Malkus}},
  \bibinfo{journal}{Proc. R. Soc. London A} \textbf{\bibinfo{volume}{225}},
  \bibinfo{pages}{196} (\bibinfo{year}{1954}).

\bibitem[{\citenamefont{Krishnamurti and Howard}(1981)}]{KH81}
\bibinfo{author}{\bibfnamefont{R.}~\bibnamefont{Krishnamurti}}
  \bibnamefont{and} \bibinfo{author}{\bibfnamefont{L.~N.}
  \bibnamefont{Howard}}, \bibinfo{journal}{Proc. Natl. Acad. Sci.}
  \textbf{\bibinfo{volume}{78}}, \bibinfo{pages}{1981} (\bibinfo{year}{1981}).

\bibitem[{\citenamefont{Brown et~al.}(2005)\citenamefont{Brown, Funfschilling,
  Nikolaenko, and Ahlers}}]{BFNA05}
\bibinfo{author}{\bibfnamefont{E.}~\bibnamefont{Brown}},
  \bibinfo{author}{\bibfnamefont{D.}~\bibnamefont{Funfschilling}},
  \bibinfo{author}{\bibfnamefont{A.}~\bibnamefont{Nikolaenko}},
  \bibnamefont{and} \bibinfo{author}{\bibfnamefont{G.}~\bibnamefont{Ahlers}},
  \bibinfo{journal}{Phys. Fluids} \textbf{\bibinfo{volume}{17}},
  \bibinfo{pages}{075108} (\bibinfo{year}{2005}).

\bibitem[{\citenamefont{Verzicco}(2004)}]{Ve04}
\bibinfo{author}{\bibfnamefont{R.}~\bibnamefont{Verzicco}},
  \bibinfo{journal}{Phys. Fluids} \textbf{\bibinfo{volume}{16}},
  \bibinfo{pages}{1965} (\bibinfo{year}{2004}).

\bibitem[{\citenamefont{Liu and Ahlers}(1997)}]{LA97}
\bibinfo{author}{\bibfnamefont{L.}~\bibnamefont{Liu}} \bibnamefont{and}
  \bibinfo{author}{\bibfnamefont{G.}~\bibnamefont{Ahlers}},
  \bibinfo{journal}{Phys. Rev. E} \textbf{\bibinfo{volume}{55}},
  \bibinfo{pages}{6950} (\bibinfo{year}{1997}).

\bibitem[{\citenamefont{Verzicco and Camussi}(2003)}]{VC03}
\bibinfo{author}{\bibfnamefont{R.}~\bibnamefont{Verzicco}} \bibnamefont{and}
  \bibinfo{author}{\bibfnamefont{R.}~\bibnamefont{Camussi}},
  \bibinfo{journal}{J. Fluid Mech.} \textbf{\bibinfo{volume}{477}},
  \bibinfo{pages}{19} (\bibinfo{year}{2003}).

\bibitem[{\citenamefont{Stringano and Verzicco}(2006)}]{SV06}
\bibinfo{author}{\bibfnamefont{G.}~\bibnamefont{Stringano}} \bibnamefont{and}
  \bibinfo{author}{\bibfnamefont{R.}~\bibnamefont{Verzicco}},
  \bibinfo{journal}{J. Fluid Mech.} \textbf{\bibinfo{volume}{548}},
  \bibinfo{pages}{1} (\bibinfo{year}{2006}).

\bibitem[{\citenamefont{Xi and Xia}(2007)}]{XX07}
\bibinfo{author}{\bibfnamefont{H.-D.} \bibnamefont{Xi}} \bibnamefont{and}
  \bibinfo{author}{\bibfnamefont{K.-Q.} \bibnamefont{Xia}},
  \bibinfo{journal}{Phys. Fluids} \textbf{\bibinfo{volume}{20}},
  \bibinfo{pages}{055104} (\bibinfo{year}{2007}).

\bibitem[{\citenamefont{Brown and Ahlers}(2007)}]{BA07_EPL}
\bibinfo{author}{\bibfnamefont{E.}~\bibnamefont{Brown}} \bibnamefont{and}
  \bibinfo{author}{\bibfnamefont{G.}~\bibnamefont{Ahlers}},
  \bibinfo{journal}{Europhys. Lett.} \textbf{\bibinfo{volume}{80}},
  \bibinfo{pages}{14001} (\bibinfo{year}{2007}).

\end{thebibliography}

\end{document}